\documentclass[aps,prl,twocolumn,groupedaddress]{revtex4}
\usepackage{graphicx}
\begin{document}

\title{Chemically Functionalized Semiconducting Carbon Nanotubes: \\
Limits for High Conductance Performance}

\author{Alejandro L\'opez-Bezanilla$^{1}$, X. Blase$^{2}$ and Stephan Roche$^{3}$,$^{4}$}
\affiliation{
$^1$ CEA, Institut of Nanosciences and Cryogenics, INAC,SPSMS,GT  17 rue des Martyrs, 
38054 Grenoble Cedex 9, France. \\
$^2$ Institut N\'{e}el, CNRS and Universit\'{e} Joseph Fourier, B.P. 166, 38042 Grenoble Cedex 09, France.\\
$^3$ CEA, Institut of Nanosciences and Cryogenics, INAC, SP2M, ${\rm L}\_{\rm Sim}$,  
17 rue des Martyrs, 38054 Grenoble Cedex 9, France. \\ 
$^4$ Institute for Materials Science, TU Dresden, D-01062 Dresden, Germany.
}

\begin{abstract}
We present a first-principles study of the electronic transport properties of micrometer long 
semiconducting CNTs randomly covered with carbene functional groups. Whereas prior 
studies suggested that metallic tubes are hardly affected by such addends, we show here that
 the conductance of semiconducting tubes with standard diameter is on the contrary severely damaged. 
The configurational averaged conductance as a function of tube diameter and with a coverage of 
up to one hundred functional groups is extracted.  Our results indicate that the search for a 
conductance-preserving covalent functionalization route remains a challenging issue.
\end{abstract}


\maketitle
\section{Introduction}
Carbon nanotubes (CNTs) have proven to be outstanding materials for exploring quantum physics 
in low dimensionality. Their unique mechanical, electronic and transport properties have  
stimulated a great amount of fundamental and applied research with potential use in various
technological devices \cite{charlier:677}. In particular, the engineering of sidewall 
chemical functionalization of CNTs has opened much perspectives in the development of innovative
systems  such as chemical sensors \cite{GuoJACS05}, optically modulated conductors \cite{SimmonsPRL07}, 
or novel switching devices and molecular memories \cite{MannikPRL06}. However, many difficulties 
remain to be overcome in order to benefit from the large diversity of organic chemistry reactions 
that can be performed on a tube wall together with the exceptional transport properties of pristine CNTs. Indeed, commonly used 
functionalization processes, such as the phenyl group addition, \cite{BharJCM01} produce a significant 
number of \textit{sp}$^3$-like defects which dramatically disrupt the $\pi$-conjugated network of CNTs,
severely damaging their conductance \cite{CabanaJACS07,LeePRL05,BezanillaNL}. To circumvent such a problem, $[2+1]$ cycloaddition reactions 
have been proposed \cite{CA}. This functionalization is driven by grafted carbene (or nitrene) groups that induce bond 
cleaving between adjacent sidewall carbon atoms, maintaining the \textit{sp}$^2$ hybridization and 
providing sites for further attachment of complex molecules and related functionalities \cite{Chen}. First-principles calculations 
have shown that high conductance is preserved in functionalized metallic nanotubes with random 
coverage of addends \cite{BezanillaNL,LeePRL06,Lu06}, but the situation for semiconducting nanotubes remains 
unexplored. Since standard devices such as field-effect transistors exclusively rely on the use of semiconducting
tubes, it is crucial to clarify the effect of functionalization of such systems.

In this paper, we report a first-principles computational study of the effect of ${\rm CH}_{2}$ 
carbene functional groups grafting on the charge transport properties of micrometer long semiconducting carbon nanotubes.
The evolution of conductance as a function of the tube diameter and with a random distributions of up to one 
hundred functional groups is explored. It is shown that for zigzag tubes, the transition from a quasi-ballistic 
to a strongly diffusive regime occurs at a much smaller tube diameter as compared to armchair metallic tubes. 
This is related to an earlier transition from the ``open"  to ``close" configuration for the nanotube
C-C bond bridged by the carbene group. As a result, for tube diameters above 1.2-1.4 nm, the conductance of 
functionalized semiconducting tubes is found to decay rapidly with increasing number of grafted groups,
in great contrast with armchair tubes of equivalent diameter.

\section{Technical details}
The interaction of functional groups with (n,0) zigzag CNTs (n stands for the nanotube index) were explored with 
the {\sc Siesta} \cite{siesta1, siesta2} code, which implements the density functional theory with a linear 
combination of numerical atomic-like orbital basis sets for describing the eigenstates. Our calculations 
were performed using the local density approximation for the exchange-correlation potential. 
Standard norm-conserving pseudopotentials \cite{Troullier} describe the interaction between ionic cores and 
valence electrons. Split-valence double-$\zeta$ polarized basis sets were used in all simulations.
Convergency tests showed that such a basis provide well converged conductance profiles.  Interactions between 
addends and their replica in the periodic one-dimensional cells were avoided by considering six-unit CNT-cells 
(25.55 \AA ) which guarantees a convergence to pristine CNT in between the image molecules.

Relaxed atomic positions are obtained when residual atomic forces become smaller than 0.04 eV/\AA.

\section{Diameter dependent functionalization induces bond cleaving}
In the case of zigzag CNTs, there are two types of C-C bond orientations onto which a ${\rm CH}_{2}$ can be grafted, namely parallel ($||$) or skewed ($\backslash$)  with respect to the tube axis. These bond orientations are depicted in Fig.~1 (inset) and Fig.~2 (left inset) respectively, where the bridged tube C-C bonds upon the carbene adsorption have been outlined (yellow links). This is at odds with the case of metallic armchair tubes for which (a) the parallel configuration does not exist, and (b) bonds perpendicular to the tube axis offer preferential addition sites \cite{LeePRL06}. In each case, the equilibrium position is reached when the carbon atom of ${\rm CH}_{2}$ is located on top of the C-C tube bond forming a bridge-like structure \cite{BezanillaNL,LeePRL06,Lu06,Lee}. Our calculations shows that the ($\backslash$) conformation is always more stable than the ($||$) one. This is consistent with the result of Chen {\it et al.} \cite{Chen} for the case of a small (8,0) finite-size cluster. It is further consistent with the study by Lee and Marzari of metallic armchair tubes showing that bonds orthogonal to the tube axis are  preferentially grafted \cite{LeePRL06,Lee}. As a rule of thumb summarizing all these results, bonds with largest angle with respect to the tube axis are more active towards adsorption, since they experience most of the tube curvature.
We therefore focus on the skewed configurations and show in Fig.~1 the evolutions of the sidewall bond distance ($d_{\rm C-C}$) between the two carbon atoms anchoring the carbene group as a function of zigzag CNT index, for both large and small gap CNTs \cite{note}. s
As observed in the case of armchair tubes for the most stable orthogonal configuration \cite{LeePRL06,Lee}, there is a transition from an ''open-skewed'' configuration at small diameter, with  $d_{\rm C-C}=2.1~$\AA \ for the (8,0) CNT, to a closed one, with $d_{\rm C-C}=1.65~$\AA \ for the (22,0) tube. The transition occurs for the (15,0) tube with a diameter of $11.7$\AA, at much smaller a diameter than for armchair tubes for which the transition occurs for the (18,18) tube (diameter $=24.5~$\AA). Such results can be rationalized by remarking that bonds along the circumferential direction undergo the largest tensile stress induced by curvature and are thus more prone to open. As a matter of fact, we find that bonds parallel to the tube axis, which do not feel the effect of curvature, always exhibit the closed configuration. An interesting observation is that large gap zigzag tubes exhibit a transition to the closed configuration at a larger diameter than small gap ones. This transition to closed configurations at rather small diameters for semiconducting tubes bears important consequences on transport, as explained hereafter. 

\section{Quantum transport in functionalized nanotubes}
To explore mesoscopic transport in long and disordered nanotubes, we follow a well established computational 
strategy as discussed in prior works \cite{Avriller06,G-N,rocha,Markussen:PRB74}. 
From the first-principles calculations for long tube sections functionalized by single groups, {\it ab initio} 
Hamiltonian and overlap matrices are obtained. The atomic-like features of the basis set utilized by SIESTA allows us to have a relatively small and manageable description of the interaction between atoms in the shape of sparse Hamiltonian matrices. Such a set of initial ``building blocks" is then used to construct  micrometer long nanotubes composed by a random serie of functionalized and pristine tube sections, introducing rotational and translational disorder. Functionalised CNT unit cells are long enough to avoid spurious interactions between virtual images during DFT-calculations, so that slab-extremes are converged to the clean system. Functionalized and clean sections of CNTs can then be matched to build up long systems with perfect contact areas between the building blocks. \cite{meth}
Standard techniques to calculate the Green's function associated with the sparse Hamiltonian and overlap matrices are further employed to include recursively the contribution of additional sections, leading to an O(N) scheme with respect to the tube length. First-principles transport calculations of complex and disordered systems with several thousands of atoms can therefore be achieved. Assuming that the total system is made by a functionalized nanotube connected to two semi-infinite perfect CNTs leads, the conductance can be computed using the Landauer-B\"uttiker transport theory 
\cite{LeePRL05}.

\begin{figure}
\includegraphics[width=0.45 \textwidth]{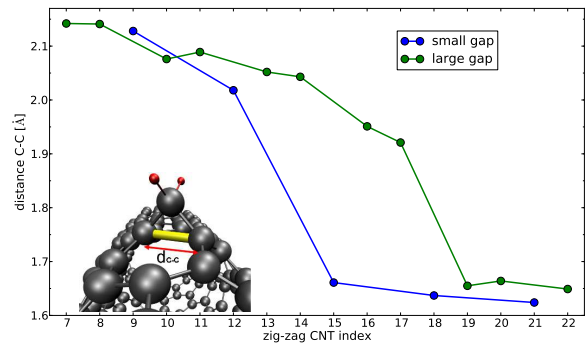}
\caption{\label{figure1} (color online) Main frame:  Sidewall equilibrium distance for C-C bond anchoring the 
carbene group, as a function of the tube (n,0) index, for both types of zigzag tubes: small gap (n=3m) and 
large gap (n=3m$\pm$1) CNTs. A ball-and-stick model illustrates the ($\backslash$) skewed bonding 
configuration.} 
\end{figure}

We first discuss the effect of a single functional group and show in Fig.~2 the energy-resolved conductance
for a (19,0) CNT (main frame) and a (12,0) CNT (inset). These tubes are on both sides of the ``open-to-close"
C-C bond distance transition for the skewed ($\backslash$) configuration (see Fig.~1). In the case of the 
small gap (12,0) CNT, the grafted group in the ($\backslash$) configuration, with open  C-C bond, has hardly any effect on the conductance as shown in the inset (blue curve). On the contrary, the parallel ($||$) bonding geometry, with
closed C-C bond, yields a dramatic drop of conductance at nearly all energies (red curve inset). Such a contrasting
behavior can be rationalized by observing that in the open geometry, the carbon atoms remain three-fold coordinated
preserving thus the conjugated character of the \textit{sp}$^2$  pristine tube network.  
\cite{LeePRL06,Lu06,BezanillaNL}  In constrast, in the closed geometry, 
the bridged carbon atoms are four-fold coordinated, destroying the local conjugated character of the
$\pi$-network and thus reinforcing backscattering probability. \cite{LeePRL05,BezanillaNL} Such a transition to a local \textit{sp}$^3$ character can be strikingly evidenced by comparing our data with the conductance profile generated by two
hydrogen atoms bonded to neighbouring C atoms on a ($||$) bond (see Fig.~2). Clearly, the two 
kinds of functionalization lead to very similar conductance profiles with up to one full conducting channel ($G_0=2e^2/h$) quenched at selected energies on the first plateaus on each side of the gap.  

Assuming that the functionalization process follows the thermodynamic limit, with all addends in the skewed
configuration, one may possibly conclude that the conductance of the small gap (12,0) tube can be preserved up to 
very large coverages. It was shown in the case of a standard (10,10) armchair tubes that up to several hundreds
of carbene groups could be grafted while preserving nearly 75$\%$ of the conductance at all energies 
\cite{BezanillaNL}. However, as shown above, the open-to-closed transition in zigzag tubes occurs 
at much smaller diameter as compared to armchair tubes, so that basically all zigzag 
tubes with a diameter larger than $\sim$ 1.2-1.5 nanometers will exhibit closed grafting configurations, 
even for the most stable skewed bonds. As a result, and as shown in Fig.~2 for the (19,0) tube, 
even an isolated carbene group grafted on a skewed bond induces severe backscattering 
close to band edges (onsets of plateaus). Even though the effect is less pronounced than that induced
by a carbene or two hydrogen atoms on a parallel bond (red and dashed green lines), the 
consequences are now shown to become rapidly dramatic for a few tens of attached functional groups.

\begin{figure}
\includegraphics[width=0.465 \textwidth]{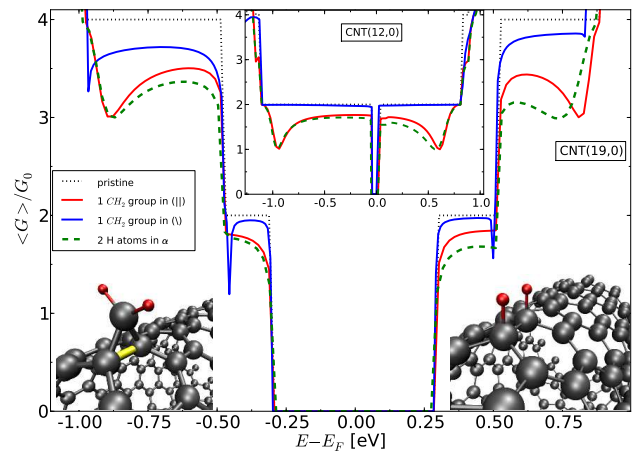}
\caption{\label{figure3}(color online) Main frame: conductance (in $G_{0}=2e^{2}/h$ unit) of pristine (19,0) tube (dotted line) and with a 
carbene grafted in the parallel ($||$) (red line) and skewed ($\backslash$) (blue line) configuration. Carbene 
in the $||$ conformation shows similar conductance profile than two grafted hydrogens (green dashed). 
Inset: Same as in main frame but for a small gap (12,0) tube. Ball-and-stick models of ($||$) 
configuration and dihydrogenation are also given in insets for illustration.}
\end{figure}

To further explore the transport properties of long and chemically modified semiconducting nanotubes, we 
have performed a mesoscopic study for several tubes with a length of 1 $\mu$m and random grafting of an increasing
number of carbene groups. Conductances were averaged over up to 20 configurations of disorder for a given tube
and number of functional groups. As a representative case, Fig.~3 shows a comparison between the (10,0) and the (22,0) tubes for a single grafted carbene and for an increasing number of groups (25, 50, and 100) in the ($\backslash$) conformation. For a single grafted group, we revisit the results of Fig.~2 established for the (12,0) and (19,0) tubes: the backscattering is negligible for very small tubes, but becomes significant for tubes such as the (22,0) with a diameter beyond the open-closed bond transition value ($\sim$ 1.2-1.4 nm). This initial difference is then amplified when adding a much larger number of addends, which enhances disorder effects and allows quantum interferences between scattering centers to set in. As a result, for the same number of grafted groups, their impact on transport in both semiconducting tubes becomes markedly different. 
While the (10,0) manifests a weak reduction of conductance on the first plateau
(which remains close to $2G_{0}$ even for the largest number of addends), a strong decay is observed for 
the (22,0) with an almost fully suppressed conductance even for the lower studied density (25 groups randomly 
distributed over 1 $\mu$m). 

\begin{figure}
\includegraphics[width=0.48 \textwidth]{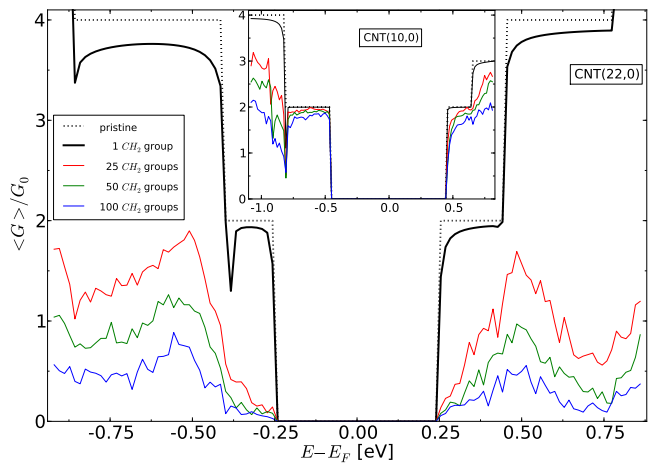}
\caption{\label{figure2} (color online) Main frame: Conductance for pristine (22,0) nanotube (dotted line), with 
a single grafted carbene in the ($\backslash$) conformation (solid line), and for averaged configurations 
with varying addends density (25, 50 or 100 ${\rm CH}_{2}$ groups distributed at random along one $\mu$m). 
Inset: Same as in main frame but for the smaller diameter (10,0) semiconducting tube.}
\end{figure}

To characterize the related conduction regimes, we report in Fig.~4, the average conductance for (12,0) 
(left frame) and (19,0) (right frame) ${\rm CH}_{2}$-functionalized CNTs as a function of the number of attached 
carbene. Both ($\backslash$) and ($||$) bonding configurations of ${\rm CH}_{2}$ are considered, together with
two-hydrogen atoms in the parallel configuration (see above). The average is performed over 200 configurations 
of disorder and at the selected energy $E-E_{F}=0.35 eV$, that is roughly at the center of the first electron
subband. For the ($\backslash$) bonding in the (12,0) tube, 
a weak conductance decay evidences a quasiballistic regime for the considered addends density. In contrast, 
for the ($||$) bonding configuration, the conducance decrease is roughly inversely 
proportional to the number of grafted groups, indicating a diffusive regime. This results in a conductance 
reduction by one order of magnitude for only 40 groups distributed over $1\mu$m, showing the large impact 
of such bonding type. The increase in diameter further degrades the conductance properties. In the (19,0)
case, even the more favorable ($\backslash$) bonding leads now to a quick drop of conductance with increasing
number of grafted functional groups. 

\begin{figure}
\includegraphics[width=0.48 \textwidth]{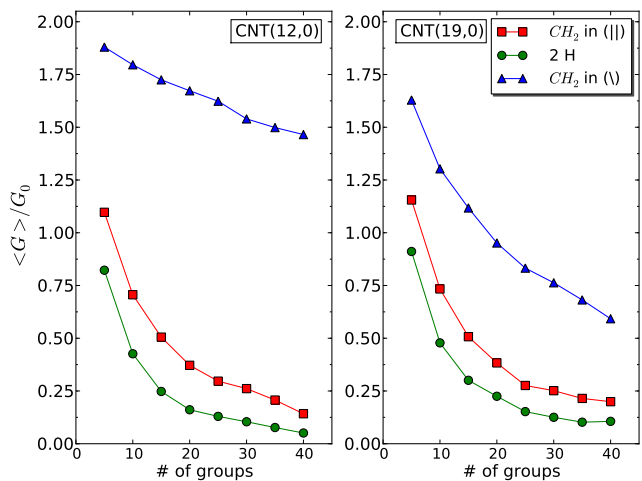}
\caption{\label{figure4} (color online)  Average conductances versus number of attached carbene groups
for the (12,0) and (19,0) nanotubes, and different conformations of carbene groups bonding.} 
\end{figure}

The early transition from the open to closed configuration for carbene groups grafted onto zigzag nanotubes, with
a critical diameter of about 1.2 nm, represents a serious drawback for the use of such functional groups
in the case of semiconducting nanotubes. While smaller diameter tubes can always be synthesized, mass produced
CNTs exhibit a mean diameter which does not go below such a critical value. As demonstrated above, for standard
semiconducting tubes, carbene group functionalization is thus expected to lead to a drastic reduction of conductance
with only a few functional groups. Besides, it was shown that the carbene group desorbs easily at room temperature
from graphene, that is in the closed C-C bond configuration \cite{roxana}. Namely, the desorption barrier (a few 
tenths of an eV) can be easily overcome in ambient conditions. It is therefore not even clear that, beyond the 
transition diameter, carbene grafting can be a viable way for functionalizing zigzag nanotubes.

\section{Conclusion}
By using a fully {\it ab initio} transport approach, we have explored the transport regimes in chemically functionalized semiconducting carbon nanotubes, assessing the true potential of cycloaddition reactions as a precursor for further enhancement of devices functionalities. Small diameter functionalized tubes  have been shown to preserve good conduction ability (as for the case of weak physisorption \cite{Latil}), whereas a strongly diffusive regime was obtained for larger diameter tubes, similar to the effect of a low dihydrogenation density. This effect was explained by the orbital rehybridization of the atoms to which functional groups are attached. Our study quantifies the severe limits of divalent additions in large diameter semiconducting nanotubes, and suggest that nanotubes with smaller diameter sizes should be targeted to engineer efficient novel device functionalities. 

\section{Acknowledgements}
We thank the CEA/CCRT supercomputing facilities for providing computational resources and technical support.  Financial support from the ANR/PNANO project ACCENT is acknowledged. A.L.B acknowledges support from the Marie-Curie fellowship CHEMTRONICS program. S.R. is indebted to the Alexander von Humboldt Foundation for financial support. Luigi Genovese and Thierry Deutsch are acknowledged for fruitful discussions.



\begin{thebibliography}{10}
\bibitem{meth} As an example of the convergence to the clean system in the extremes of a functionalized nanotube we show with dotted lines in Fig.~2 and Fig.~3 the transport calculations on pristine CNTs, the energetic descriptions of which correspond to sub-Hamiltonian od the functinalized CNT extreme layers.
\end{thebibliography}
\end{document}